\documentclass[aps,pra,twocolumn,showpacs]{revtex4-1}

\usepackage{amsmath,amssymb}
\usepackage{graphicx}

\usepackage{color}

\graphicspath{ {.} {./pics/} }

\newcommand{\rmi}{\mathrm{i}}
\newcommand{\rmd}{\mathrm{d}}
\newcommand{\rme}{\mathrm{e}}
\newcommand{\ti}{i}
\newcommand{\tp}{p}
\newcommand{\ts}{s}

\begin{document}

\title{Entangled light from driven dissipative microcavities}

\author{D. Pagel}
\email{pagel@physik.uni-greifswald.de}
\affiliation{Institut f\"ur Physik, Ernst-Moritz-Arndt-Universit\"at Greifswald, 17487 Greifswald, Germany}
\author{H. Fehske}
\affiliation{Institut f\"ur Physik, Ernst-Moritz-Arndt-Universit\"at Greifswald, 17487 Greifswald, Germany}

\begin{abstract}
We study the generation of entangled light in planar semiconductor microcavities.
The focus is on a particular pump configuration where the dissipative internal polariton dynamics leads to the emission of entangled light in a W-state.
Our study is based on the nonlinear equations of motion for the polariton operators derived within the dynamics-controlled truncation formalism.
They include the losses through the cavity mirrors, the interaction with lattice vibrations, and the external laser driving in a Langevin approach.
We find that the generated entanglement is robust against decoherence under realistic experimental conditions.
Our results show that pair correlations in solid-state devices can be used to stabilize the nonlocal properties of the emitted radiation.
\end{abstract}


\maketitle

\section{Introduction}
Quantum entanglement is known as the essential resource for quantum information processing\ \cite{HHHH09, GT09, NC10}.
It is defined as a nonlocal correlation that cannot be interpreted in terms of classical joint probabilities\ \cite{EPR35, Sch35, Wer89}.
The most fundamental examples of nonequivalent forms are GHZ and W states\ \cite{GHZ89, DVC00}.
The identification and quantification of entanglement is commonly based on entanglement witnesses\ \cite{HHH96, HHH01, Bra05, SV11a, SV13, RSV15}. The generation and control of entangled states still is a challenging task for quantum computation.
In the optical domain, the implementation of quantum algorithms relies on the availability of efficient sources for entangled photons.

The generation of entangled photons is usually based on parametric down conversion in nonlinear crystals\ \cite{KMWZSS95, AK04} or biexciton decay in quantum dots\ \cite{BSPY00, HSK03}. Optically excited semiconductor microcavities\ \cite{WNIA92, HWSOPI94, Lang04, CBC05, Ciu04} are alternative candidates for the efficient generation of entangled light on the $\mu$m scale\ \cite{Ciu04, CSQ01, SDSSL05, PDSSS09, PEVWR14, EVWP15}.
The exciting laser field with frequency near the fundamental band gap coherently generates electron-hole pairs (excitons).
The dynamical evolution of excitons is governed by the Coulomb interaction, and the efficient coupling to the cavity photons leads to mixed exciton-photon modes---so-called polaritons\ \cite{Hop58, OSS98, TY99}.
The external laser can be tuned to stimulate parametric scattering processes between polaritons which may cause entanglement\ \cite{PFSV12, PFSV13}.
A moving polariton induces an electric polarization as a source of light that carries the initial (internal) polariton entanglement\ \cite{CBC05}.
Then, depending on the explicit pump configuration, branch or frequency entanglement\ \cite{Ciu04, PFSV12}, polarization entanglement\ \cite{EVWP15}, multipartite entanglement\ \cite{PFSV13}, and hyper-entangled photon pairs in multiple coupled microcavities\ \cite{PEVWR14} can be generated.

In this work, we demonstrate that a semiconductor microcavity can be used to entangle light in a W state configuration. Beyond that,   
such a setup allows to analyze the internal polariton entanglement properties in the presence of dissipation.
Specifically, we consider a microcavity that is either continuously driven or excited by Gaussian pump pulses.
In previous work~\cite{PFSV13},  we introduced the specific pump arrangement for the creation of entangled light in a W state 
and used multipartite entanglement witnesses to verify the nonlocal correlations.
Here, the inclusion of decoherence, induced by the losses through the cavity mirrors, and the coupling to lattice vibrations, within the dynamics-controlled truncation formalism, allows us to study the emitted light under realistic experimental conditions.
Most notably, we show that the entanglement of the generated light is robust against dephasing.

We proceed as follows.
In Sec.\ \ref{sec:eom} we briefly recapitulate the equations of motion obtained within the dynamics-controlled truncation formalism and review the explicit pump configuration.
The tomographic reconstruction of the state of the emitted radiation is performed in Sec.\ \ref{sec:res}, including the analytical solution in the limit of continuous pumping in Sec.\ \ref{ssec:ana} and the numerical solution for Gaussian pump pulses in Sec.\ \ref{ssec:num}.
Further details for the derivation of the analytical result can be found in App.\ \ref{app:sol}.
We finally conclude in Sec.\ \ref{sec:con}.

\section{\label{sec:eom}Theoretical description of parametric emission}
The theoretical description of the dynamical processes in semiconductor microcavities is frequently based on an explicit bosonization of the whole system Hamiltonian\ \cite{Usui60, CSQ01, CBC05}.
Alternatively, one can derive equations of motion for generalized Hubbard (transition) operators and truncate these equations at a certain order 
of the external field. This approach is called dynamics-controlled truncation scheme\ \cite{AS94, SG96, PDSSRG08a, PDSSRG08b}; it can naturally be used to evaluate correlation functions that allow for a tomographic reconstruction of the state of the emitted light modes\ \cite{PDSSS09, EVWP15}.
It is thus well suited to study the generation of entangled light in semiconductor microcavities under realistic experimental conditions.

\subsection{Nonlinear system dynamics}
We begin with the equations of motion for the semiconductor exciton and cavity photon operators that are derived within the dynamics-controlled truncation formalism\ \cite{PDSSRG08a, PDSSRG08b},
\begin{subequations}\label{eom_ab}\begin{eqnarray}
  \frac{\rmd}{\rmd t} a_{\mathbf{k}} &=& -\rmi \omega_{\mathbf{k}}^c a_{\mathbf{k}} + \rmi \Omega_R b_{\mathbf{k}} \,, \\
  \frac{\rmd}{\rmd t} b_{\mathbf{k}} &=& -\rmi \omega_{\mathbf{k}}^x b_{\mathbf{k}} + \rmi \Omega_R a_{\mathbf{k}} - \rmi R_{\mathbf{k}}^{{NL}} \,,
\end{eqnarray}\end{subequations}
where $R_{\mathbf{k}}^{{NL}} = R_{\mathbf{k}}^{{sat}} + R_{\mathbf{k}}^{{xx}}$,
\begin{subequations}\begin{eqnarray}
  R_{\mathbf{k}}^{{sat}} &=& \frac{\Omega_R}{n_{{sat}}} \sum_{\mathbf{k}_1, \mathbf{k}_2} b_{\mathbf{k}_1 + \mathbf{k}_2 - \mathbf{k}}^\dagger b_{\mathbf{k}_1}^{} a_{\mathbf{k}_2}^{} \,, \\
  R_{\mathbf{k}}^{{xx}} &=& V_{xx} \sum_{\mathbf{k}_1, \mathbf{k}_2} b_{\mathbf{k}_1 + \mathbf{k}_2 - \mathbf{k}}^\dagger b_{\mathbf{k}_1}^{} b_{\mathbf{k}_2}^{} \,.
\end{eqnarray}\end{subequations}
In these equations, $a_{\mathbf{k}}$ ($b_{\mathbf{k}}$) annihilates a cavity photon (semiconductor exciton) with in-plane wave vector $\mathbf{k}$ and energy $\omega_{\mathbf{k}}^c$ ($\omega_{\mathbf{k}}^x$).
$\Omega_R$ is the dipole coupling strength between excitons and photons---the so-called Rabi frequency.
In addition, $n_{{sat}}$ denotes the exciton saturation density and $V_{xx}$ is the exciton-exciton coupling strength.

The unitary Hopfield transformation to polaritons\ \cite{Hop58},
\begin{equation}
  \begin{pmatrix} p_{1 \mathbf{k}} \\ p_{2 \mathbf{k}} \end{pmatrix} = \begin{pmatrix} X_{1 \mathbf{k}} & C_{1 \mathbf{k}} \\ X_{2 \mathbf{k}} & C_{2 \mathbf{k}} \end{pmatrix} \begin{pmatrix} b_{\mathbf{k}} \\ a_{\mathbf{k}} \end{pmatrix} \,,
\end{equation}
which is tuned to diagonalize the linear part of the equations of motion\ \eqref{eom_ab}, leads to the equations of motion in the polariton basis,
\begin{subequations}\label{eom_p}\begin{eqnarray}
  \frac{\rmd}{\rmd t} p_{1 \mathbf{k}} &=& -\rmi \omega_{1 \mathbf{k}} p_{1 \mathbf{k}} - \rmi R_{1 \mathbf{k}}^{{NL}} \,, \\
  \frac{\rmd}{\rmd t} p_{2 \mathbf{k}} &=& -\rmi \omega_{2 \mathbf{k}} p_{2 \mathbf{k}} - \rmi R_{2 \mathbf{k}}^{{NL}} \,,
\end{eqnarray}\end{subequations}
with $R_{j \mathbf{k}}^{{NL}} = X_{j \mathbf{k}} R_{\mathbf{k}}^{{NL}}$.
Here, $p_{j \mathbf{k}}$ annihilates a polariton with dispersion $\omega_{j \mathbf{k}}$ in the lower ($j = 1$) or upper ($j = 2$) branch.

\subsection{External driving and dissipation}
To include losses through the mirrors, the interaction with lattice vibrations and the external laser driving, we couple the system dynamics to the environment.
As shown in Ref.\ \cite{PDSSRG08a}, combining the dynamics-controlled truncation scheme with the nonequilibrium quantum Langevin approach, the incoherent system dynamics decouples from parametric scattering processes.
In particular, we have to add the damping rates $\Gamma_{j \mathbf{k}}^{(tot)}$ and Langevin noise source operators $\mathcal{F}$ with proper statistics and moments to the equations of motion\ \eqref{eom_p}.
This yields
\begin{subequations}\label{eom_pF}\begin{eqnarray}
  \frac{\rmd}{\rmd t} p_{1 \mathbf{k}} &=& -\rmi \widetilde{\omega}_{1 \mathbf{k}} p_{1 \mathbf{k}} - \rmi R_{1 \mathbf{k}}^{{NL}} + \mathcal{F}_{p_{1 \mathbf{k}}} \,, \\
  \frac{\rmd}{\rmd t} p_{2 \mathbf{k}} &=& -\rmi \widetilde{\omega}_{2 \mathbf{k}} p_{2 \mathbf{k}} - \rmi R_{2 \mathbf{k}}^{{NL}} + \mathcal{F}_{p_{2 \mathbf{k}}} \,,
\end{eqnarray}\end{subequations}
where $\widetilde{\omega}_{j \mathbf{k}} = \omega_{j \mathbf{k}} - \rmi \Gamma_{j \mathbf{k}}^{(tot)} / 2$.
The  operators $\mathcal{F}$ are characterized by vanishing expectation values, $\langle \mathcal{F}_\mu \rangle = 0$, where $\mu = p_{j \mathbf{k}}^{(\dagger)}$, and by the second order moments
\begin{equation}\label{FF}
  \langle \mathcal{F}_\mu(t) \mathcal{F}_\nu(t') \rangle = 2 \langle D_{\mu\nu}(t) \rangle \, \delta(t - t')
\end{equation}
with diffusion coefficients
\begin{equation}\label{D}
  2 \langle D_{\mu\nu}(t) \rangle = \frac{\rmd}{\rmd t} \langle \mu(t) \nu(t) \rangle - \langle \dot{\mu}(t) \nu(t) + \mu(t) \dot{\nu}(t) \rangle \,.
\end{equation}
In Eq.\ \eqref{D}, the dot denotes the time-derivative following from Eqs.\ \eqref{eom_p}, i.e., without  noise source operators $\mathcal{F}$.

Equations of motion for the expectation values $\langle \mu(t) \nu(t) \rangle$---the so-called polariton photoluminescence---are given in Ref.\ \cite{PDSSRG08a}.
They are derived in the framework of a second order Born-Markov approach.
Important for us is the final result: Due to the decoupling of incoherent dynamics and parametric processes the diffusion coefficients in Eq.\ \eqref{FF} can be used as input when we calculate multi-time correlation functions of polariton operators.
We stress that the damping rates $\Gamma_{j \mathbf{k}}^{(tot)}$  follow from this treatment too.

\subsection{Explicit pump scenario}
\begin{figure}
  \includegraphics[width=0.7\linewidth]{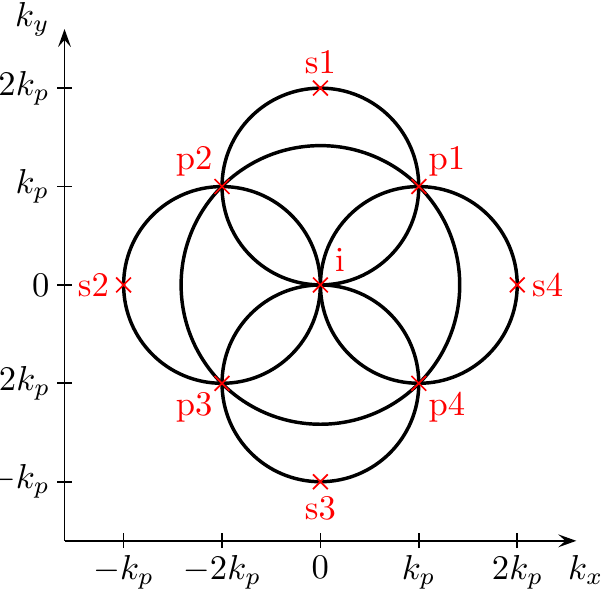}
  \caption{\label{fig:setup}%
    Sketch of the considered pump-signal-idler configuration in momentum space.
    The circles depict the possible energy- and momentum-conserving scattering processes within the lower branch, where two pumped polaritons scatter into pairs of signal/idler polaritons.
    Specifically, mixed-pump scattering processes of oppositely arranged (neighboring) pumps with $|\mathbf{k}_{\tp n} + \mathbf{k}_{\tp m}| = 0$ ($|\mathbf{k}_{\tp n} + \mathbf{k}_{\tp m}| = 2 k_\tp$) contribute to the circle(s) with radius $\sqrt{2} k_\tp$ ($k_\tp$).
  }
\end{figure}

Let us now  consider the experimental setup introduced in Ref.\ \cite{PFSV13}, where four pump lasers drive the lower polariton branch at  wave vectors $\mathbf{k}_{\tp1} = (k_\tp, k_\tp)$, $\mathbf{k}_{\tp2} = (-k_\tp, k_\tp)$, $\mathbf{k}_{\tp3} = (-k_\tp, -k_\tp)$, and $\mathbf{k}_{\tp4} = (k_\tp, -k_\tp)$ (see Fig.\ \ref{fig:setup}).
The incident angles of all pumps are below the magic angle\ \cite{SBSSWR00, Lang04} such that single-pump scattering processes (signal at $\mathbf{k}$ and idler at $2 \mathbf{k}_{\tp n} - \mathbf{k}$) are negligible.
The multi-pump parametric processes (signal at $\mathbf{k}$ and idler at $\mathbf{k}_{\tp n} + \mathbf{k}_{\tp m} - \mathbf{k}$ with $n \neq m$) share a common idler mode at $\mathbf{k}_\ti = (0, 0)$. The four corresponding signal modes at $\mathbf{k}_{\ts1} = (0, 2 k_\tp)$, $\mathbf{k}_{\ts2} = (-2 k_\tp, 0)$, $\mathbf{k}_{\ts3} = (0, -2 k_\tp)$, and $\mathbf{k}_{\ts4} = (2 k_\tp, 0)$ have  been shown to be entangled~\cite{PFSV13}.

To obtain the equations of motion for the signal and idler modes, we introduce a simplified notation.
In particular, because all scattering processes are within the lower polariton branch, we omit the branch index.
In addition, we introduce $Y_x = Y_{1 \mathbf{k}_x}$ for every quantity $Y = P, \omega, \Gamma^{(tot)}, \widetilde{\omega}, R^{{NL}}, X, C$ and define $\gamma_x = \Gamma_x^{(tot)} / 2$ for $x = \ti, \ts1, \dots, \ts4, \tp1, \dots, \tp4$.
Because of the particular pump-signal-idler configuration, we have $\omega_{\ts n} \equiv \omega_\ts$, $\gamma_{\ts n} \equiv \gamma_\ts$, $X_{\ts n} \equiv X_\ts$, $C_{\ts n} \equiv C_\ts$, $\omega_{\tp n} \equiv \omega_\tp$, $X_{\tp n} \equiv X_\tp$, and $C_{\tp n} \equiv C_\tp$ for $n = 1, \dots, 4$.

Assuming classical pump fields $\langle p_{j \mathbf{k}_{\tp n}} \rangle = \mathcal{P}_n \in \mathbb{C}$, which imply a coherent driving, and identical pumps $\mathcal{P}_n \equiv \mathcal{P}$,  we retain only terms containing the semiclassical pump amplitude $\mathcal{P}$ twice.
Introducing the vectors $\mathbf{P} = (p_\ti, p_{\ts1}^\dagger, \dots, p_{\ts4}^\dagger)^T$ and $\mathbf{F} = (F_{p_\ti}, F_{p_{\ts1}^\dagger}, \dots, F_{p_{\ts4}^\dagger})^T$, the equation of motion for the signal and idler modes takes the form:
\begin{equation}\label{eom_P}
  \frac{\rmd}{\rmd t} \mathbf{P}(t) = \mathbf{M}(t) \mathbf{P}(t) + \mathbf{F}(t)\,,
\end{equation}
where
\begin{equation}
  \mathbf{M} = \begin{pmatrix} -\rmi \widetilde{\omega}_\ti & -\rmi g_\ts \mathcal{P}^2 & -\rmi g_\ts \mathcal{P}^2 & -\rmi g_\ts \mathcal{P}^2 & -\rmi g_\ts \mathcal{P}^2 \\ \rmi g_\ts (\mathcal{P}^*)^2 & \rmi \widetilde{\omega}_\ts^* & 0 & 0 & 0 \\ \rmi g_\ts (\mathcal{P}^*)^2 & 0 & \rmi \widetilde{\omega}_\ts^* & 0 & 0 \\ \rmi g_\ts (\mathcal{P}^*)^2 & 0 & 0 & \rmi \widetilde{\omega}_\ts^* & 0 \\ \rmi g_\ts (\mathcal{P}^*)^2 & 0 & 0 & 0 & \rmi \widetilde{\omega}_\ts^* \end{pmatrix} \,,
\end{equation}
and
\begin{equation}
  g_\ts = 2 X_\ti X_\ts X_\tp \left( \frac{\Omega_R}{n_{{sat}}} C_\tp + V_{xx} X_\tp \right) \,.
\end{equation}
We note that the matrix $\mathbf{M}$ depends on time solely, because the pump amplitude $\mathcal{P}$ is time-dependent.

\subsection{State of emitted field}
Defining the matrix $\mathbf{G}$ of Green functions as the solution of the homogeneous equation,
\begin{equation}\label{green}
  \frac{\rmd}{\rmd t} \mathbf{G}(t, t') = \mathbf{M}(t) \mathbf{G}(t, t')
\end{equation}
with initial condition $\mathbf{G}(t, t) = \mathbb{I}$ ($\mathbb{I}$ is the 5$\times$5 identity matrix), the solution of the inhomogeneous Eq.\ \eqref{eom_P} is
\begin{equation}
  \mathbf{P}(t) = \mathbf{G}(t, 0) \mathbf{P}(0) + \int_0^t \mathbf{G}(t, \tau) \mathbf{F}(\tau) \: \rmd\tau \,.
\end{equation}
It allows for the calculation of multi-time correlation functions.

As a basis for the tomographic reconstruction of the measured signal/idler photon density matrix we choose the four states $|1_\ti, 1_{\ts n}\rangle$ ($n = 1, \dots, 4$), where $|1_x\rangle$ denotes the state of a photon in channel $x = \ti, \ts1, \dots, \ts4$.
This choice can experimentally be realized by postselection of events, where a click in the idler detector occurs, which takes out the vacuum component.
Then the matrix elements of the measured photon density matrix $\rho_{\ti, \ts m; \ti, \ts n} = \langle 1_\ti, 1_{\ts m} | \rho | 1_\ti, 1_{\ts n} \rangle$ are given by
\begin{widetext}\begin{eqnarray}
 \rho_{\ti, \ts m; \ti, \ts n} &=& \frac{1}{\mathcal{N}} \int_{T_d} \int_{T_d} \langle p_\ti^\dagger(t_1) p_{\ts m}^\dagger(t_2) p_{\ts n}(t_2) p_\ti(t_1) \rangle \, \rmd t_1 \, \rmd t_2 \nonumber\\
  &=& \frac{1}{\mathcal{N}} \int_{T_d} \int_{T_d} \Big[ \langle p_\ti^\dagger(t_1) p_\ti(t_1) \rangle \langle p_{\ts m}^\dagger(t_2) p_{\ts n}(t_2) \rangle + \langle p_\ti^\dagger(t_1) p_{\ts m}^\dagger(t_2) \rangle \langle p_{\ts n}(t_2) p_\ti(t_1) \rangle \Big] \, \rmd t_1 \, \rmd t_2 \,,
\end{eqnarray}\end{widetext}
where the second line follows from a Wick factorization.
In this equation $\mathcal{N}$ is a normalization constant and $T_d$ is the detector window.

\section{\label{sec:res}Results}
The equations from the last section allow us to study the tomographic reconstruction of the state of the emitted signal/idler fields in different situations.
If the semiconductor microcavity is continuously pumped, the equations of motion can be solved analytically through transformation into the pump rotating frame.
For Gaussian pump pulses---the usual experimental situation---the equations have to be solved numerically.

\subsection{\label{ssec:ana}Analytical modeling}
To obtain analytical results for the stationary state in the long-time limit we assume a continuous pumping, i.e., $\mathcal{P} = \overline{\mathcal{P}} \rme^{-\rmi \omega_\tp t}$ with $\overline{\mathcal{P}} \in \mathbb{R}$.
We define $\Delta = g_\ts \overline{\mathcal{P}}^2$ for abbreviation and perform a transformation into the pump rotating frame:
\begin{equation}
  \overline{p}_{\ts n}^\dagger = p_{\ts n}^\dagger \, \rme^{-2 \rmi \omega_\tp t} \,, \qquad
  \overline{F}_{p_{\ts n}^\dagger} = F_{p_{\ts n}^\dagger} \, \rme^{-2 \rmi \omega_\tp t} \,.
\end{equation}
Defining
\begin{equation}
  \mathbf{T}(t) = \begin{pmatrix} 1 & 0 & 0 & 0 & 0 \\ 0 & \rme^{-2 \rmi \omega_\tp t} & 0 & 0 & 0 \\ 0 & 0 & \rme^{-2 \rmi \omega_\tp t} & 0 & 0 \\ 0 & 0 & 0 & \rme^{-2 \rmi \omega_\tp t} & 0 \\ 0 & 0 & 0 & 0 & \rme^{-2 \rmi \omega_\tp t} \end{pmatrix} \,,
\end{equation}
and
\begin{equation}
   \overline{\mathbf{P}}(t) = \mathbf{T}(t) \mathbf{P}(t) \,, \qquad
 \overline{\mathbf{F}}(t) = \mathbf{T}(t) \mathbf{F}(t) 
\end{equation}
brings the equation of motion\ \eqref{eom_P} to the form
\begin{equation}
  \frac{\rmd}{\rmd t} \overline{\mathbf{P}}(t) = \overline{\mathbf{M}} \, \overline{\mathbf{P}}(t) + \overline{\mathbf{F}}(t) \,,
\end{equation}
with the time-independent matrix
\begin{eqnarray}\label{Mbar}
  \overline{\mathbf{M}} &=& \mathbf{T}^{-1}(t) \mathbf{M}(t) \mathbf{T}(t) - 2 \rmi \omega_\tp \begin{pmatrix} 0 & 0 & 0 & 0 & 0 \\ 0 & 1 & 0 & 0 & 0 \\ 0 & 0 & 1 & 0 & 0 \\ 0 & 0 & 0 & 1 & 0 \\ 0 & 0 & 0 & 0 & 1 \end{pmatrix} \nonumber\\
  &=& \begin{pmatrix} -\gamma_\ti & -\rmi \Delta & -\rmi \Delta & -\rmi \Delta & -\rmi \Delta \\ \rmi \Delta & -\gamma_\ts & 0 & 0 & 0 \\ \rmi \Delta & 0 & -\gamma_\ts & 0 & 0 \\ \rmi \Delta & 0 & 0 & -\gamma_\ts & 0 \\ \rmi \Delta & 0 & 0 & 0 & -\gamma_\ts \end{pmatrix} - \rmi \omega_\ti \mathbb{I} \,.
\end{eqnarray}
According to $\overline{\mathbf{G}}(t, t') = \mathbf{T}^{-1}(t) \mathbf{G}(t, t') \mathbf{T}(t')$, the Green functions become a matrix exponential
\begin{equation}
  \overline{\mathbf{G}}(t, t') = \exp\{\overline{\mathbf{M}} (t - t')\} = \overline{\mathbf{G}}(t - t') \,.
\end{equation}

In order to calculate the populations and correlators in the long-time limit needed for the tomographic state reconstruction, we assume  the condition $\gamma_\ts \gamma_\ti > 4 \Delta^2$ to be fulfilled.
This guarantees that all eigenvalues of  $\overline{\mathbf{M}}$ have negative real parts, i.e.,  the corresponding Green functions converge in the long-time limit. In addition, we assume a uniform noise background $N_b$, characterized by  $\langle \mathcal{F}_{p_x}(t) \mathcal{F}_{p_y}(t') \rangle = \langle \mathcal{F}_{p_x^\dagger}(t) \mathcal{F}_{p_y^\dagger}(t') \rangle = 0$,  $\langle \mathcal{F}_{p_x^\dagger}(t) \mathcal{F}_{p_y}(t') \rangle = N_b \Gamma_x \delta_{x,y} \delta(t - t')$, and $\langle \mathcal{F}_{p_x}(t) \mathcal{F}_{p_y^\dagger}(t') \rangle = (N_b + 1) \Gamma_x \delta_{x,y} \delta(t - t')$ with $x, y = \ti, \ts1, \dots, \ts4$.
As shown in App.\ \ref{app:sol},  the  tomographic reconstruction  is
\begin{equation}\label{rho}
 \rho = \frac{X}{4} \begin{pmatrix} 1 & 1 & 1 & 1 \\ 1 & 1 & 1 & 1 \\ 1 & 1 & 1 & 1 \\ 1 & 1 & 1 & 1 \end{pmatrix} + \frac{1 - X}{4} \begin{pmatrix} 1 & 0 & 0 & 0 \\ 0 & 1 & 0 & 0 \\ 0 & 0 & 1 & 0 \\ 0 & 0 & 0 & 1 \end{pmatrix} 
\end{equation}
with $X \in [0, 1]$. 
The state $\rho$ in Eq.\ \eqref{rho} is a mixture of a pure (fully entangled) W state and a (not entangled) identity state, where 
the parameter $X$ is the weight of the fully entangled W state in the mixture. 
For $X = 1$, $\rho$ is fully entangled. Contrariwise, $\rho$ is fully separable for $X = 0$. Clearly the state $\rho$ is entangled for any finite $X > 0$.  
In this sense, $X$ can be taken as an entanglement measure, which quantifies the violation of a corresponding Bell inequality.

\begin{figure}
  \includegraphics[width=0.9\linewidth]{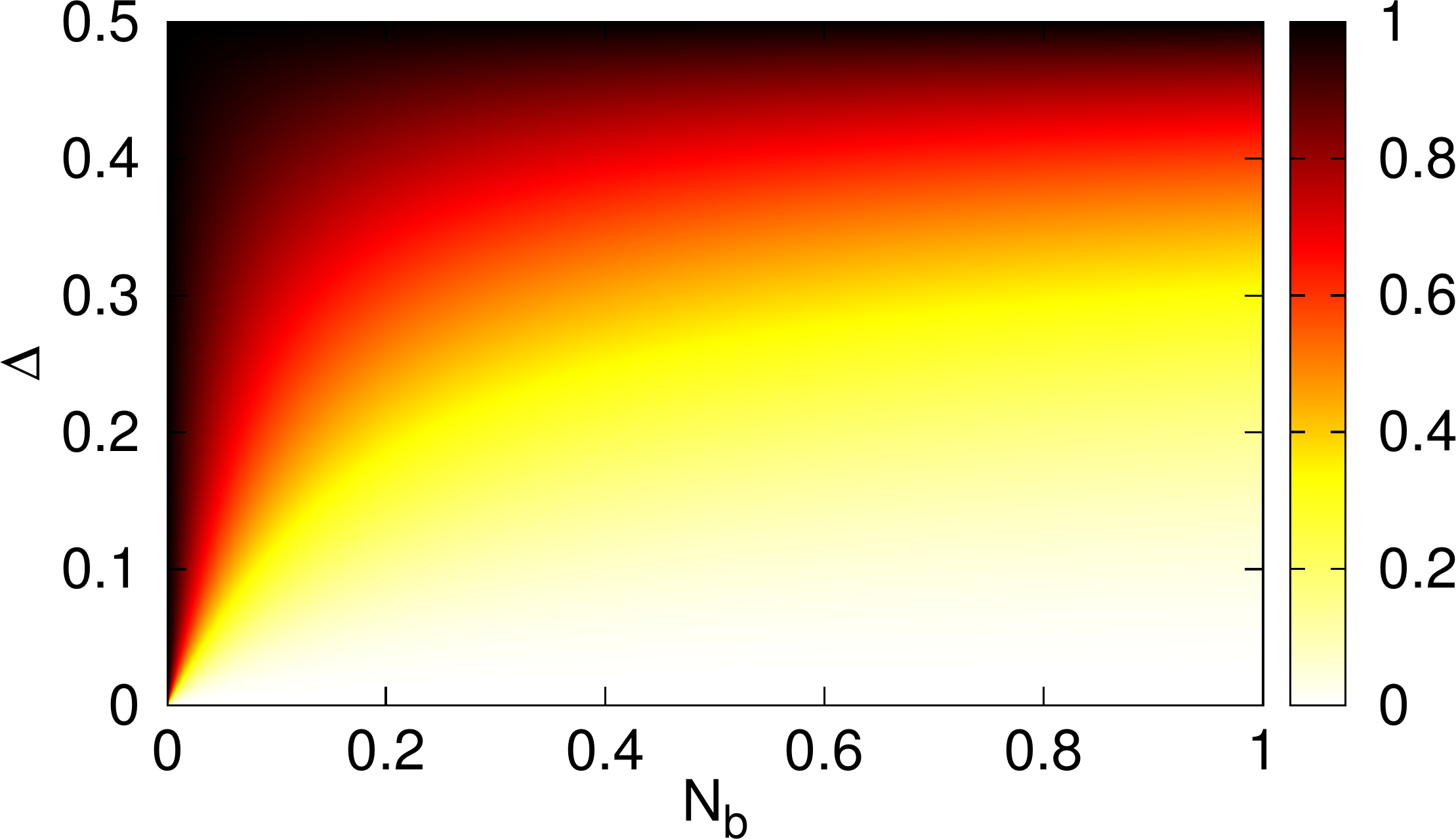}
  \caption{\label{fig:ent_2D}%
    Amount of entanglement in the state $\rho$ [Eq.\ \eqref{rho}], quantified by $X$ as a function of $\Delta$ and $N_b$, for  $\gamma_\ti = \gamma_\ts = 1$.}
\end{figure}

Figure\ \ref{fig:ent_2D} shows $X$ as a function of $\Delta$ and $N_b$.
We note that the parameter $\Delta$ is proportional to the pump intensity, which, however, is limited by the stationarity condition $4 \Delta^2 < \gamma_\ti \gamma_\ts$. Obviously, the fully entangled pure W state is obtained for vanishing noise background.
This result is in accordance with the discussion in our previous article\ \cite{PFSV13}, where losses through the cavity mirrors and the coupling to lattice vibrations are neglected. Interestingly, even for a finite noise background $N_b > 0$ 
the pure W state can be generated if the pump power is high enough. Lowering the pump power at fixed $N_b$ leads to a decrease of entanglement.

\subsection{\label{ssec:num}Numerical solution}
\begin{figure}
  \includegraphics[width=0.9\linewidth]{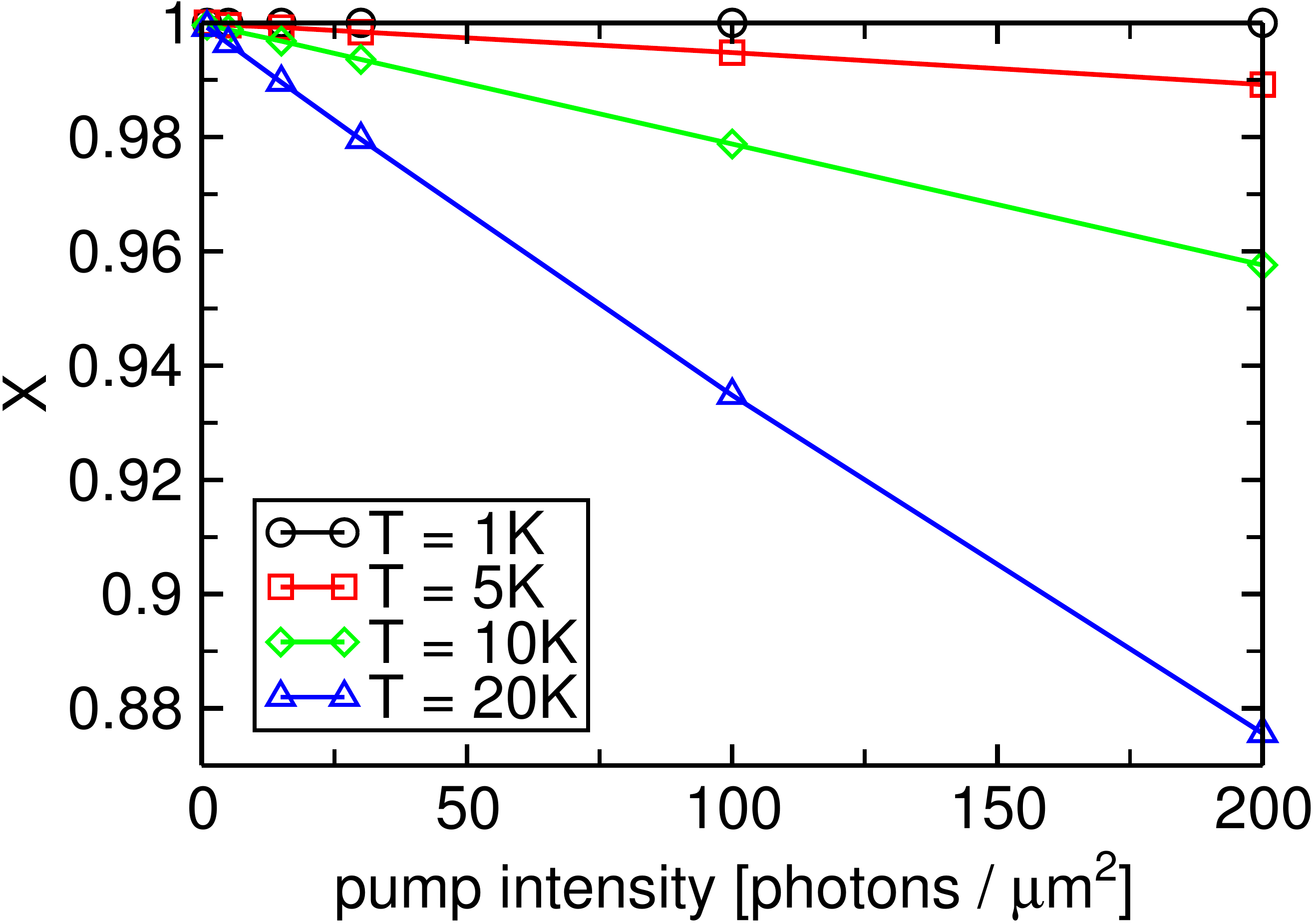}
  \caption{\label{fig:ent_num}%
    Entanglement of the numerically reconstructed signal density matrix as a function of the pump intensity for different environment temperatures.}
\end{figure}

Numerically, we can also study the case of Gaussian pump pulses.
In practice, we  solve Eq.\ \eqref{green} for Gaussian pump pulses, having an intensity maximum at 4\,ps and a width of 1\,ps, and calculate the populations and correlators to do the tomography. Thereby we choose a reasonable detection window of $T_d = 120$\,ps, allowing for technically  feasible experiments with standard photodetectors.
Again, the tomographic reconstruction results in a state of the form\ \eqref{rho}, i.e., it is fully characterized by a single parameter $X$.
The results for the amount of entanglement quantified by $X$ as a function of the pump intensity for different environment temperatures are shown in Fig.\ \ref{fig:ent_num}.
Compared to the analytical solution in the last section, we have set the uniform noise background $N_b$ to 0.
Noise enters the equations through the pump-induced photoluminescence\ \cite{RMTKSM95} that depends on the temperature of the reservoirs.
The choice $N_b = 0$ is the reason why the numerical results tend to 1 for vanishing pump intensity.
Contrary to the long-time behavior for continuous pumps, an increase in the pump intensity leads to a decrease of the entanglement.
This behavior is even more pronounced for higher reservoir temperatures.
The reason for this is a temperature-dependent background, created by the pump-induced photoluminescence, on top of which parametric scattering, i.e., entanglement generation, takes place.
Increasing the temperature at fixed pump intensity leads to a higher background at a fixed number of parametric scattering processes and hence to a lower degree of entanglement.
Nevertheless, the generated entanglement is surprisingly robust (see the range of $X$ in Fig.\ \ref{fig:ent_num}), even in the full simulation which includes the losses through the cavity mirrors and the coupling to lattice vibrations.

\section{\label{sec:con}Conclusions}
We have studied the generation of multipartite entangled light in semiconductor microcavities within the dynamics-controlled truncation scheme.
Including the losses through the cavity mirrors and the coupling to lattice vibrations, this formalism allows for a decoupling of the incoherent system dynamics (pump-induced photoluminescence) from the parametric scattering processes as the source of entanglement.
Calculating particular multi-time correlations functions, the state of the emitted signal/idler fields is obtained through tomographic reconstruction.
The resulting multipartite entanglement between the four signal channels for both continuous pumping and Gaussian pump pulses is robust against decoherence under realistic experimental conditions.
This observation shows that the emitted photons carry the initial polariton entanglement.
Since polaritons are quasiparticles composed of cavity photons and semiconductor excitons, they can sustain pair correlations over long times and distances inside such solid-state devices.
In this sense, the emitted photons serve as a probe of the internal entanglement properties.\\[1cm]


\acknowledgments
The authors would like to thank Z.~V\"{o}r\"{o}s and S.~Portolan for valuable discussions;
D.P. acknowledges the hospitality at the Photonics group at the University of Innsbruck.
This work was supported by Deutsche Forschungsgemeinschaft (Germany) through SFB 652, project B5.

\appendix
\begin{widetext}

\section{\label{app:sol}Explicit solution for continuous pumping}
Here, we evaluate the stationary populations and correlations in the long-time limit.
We start with the diagonalization of the matrix $\overline{\mathbf{M}}$ from Eq.\ \eqref{Mbar}.
The eigenvalues of $\overline{\mathbf{M}}$ are
\begin{equation}
  \lambda_1 = \lambda_2 = \lambda_3 = -\gamma_\ts - \rmi \omega_\ti \;, \qquad
  \lambda_{4/5} = -\frac{1}{2} \left( \gamma_\ti + \gamma_\ts \pm \sqrt{(\gamma_\ti - \gamma_\ts)^2 + 16 \Delta^2} \right) - \rmi \omega_\ti \,.
\end{equation}
To simplify the notation, we introduce $\Omega = \sqrt{(\gamma_\ti - \gamma_\ts)^2 + 16 \Delta^2}$, $\lambda = -\gamma_\ts - \rmi \omega_\ti$, and $\lambda_\pm = \lambda - \lambda_{4/5} = (\gamma_\ti - \gamma_\ts \pm \Omega) / 2$ with $\lambda_+ > 0$ and $\lambda_- < 0$.
With these definitions, the matrix $\mathbf{V}$ of eigenvectors is
\begin{equation}
  \mathbf{V} = \frac{1}{\sqrt{2 \Omega \lambda_+ |\lambda_-|}} \begin{pmatrix} 0 & 0 & 0 & -\sqrt{2 |\lambda_-|} \lambda_+ & \sqrt{2 \lambda_+} |\lambda_-| \\ -\sqrt{\Omega \lambda_+ |\lambda_-|} & -\sqrt{\Omega \lambda_+ |\lambda_-|} & -\sqrt{\Omega \lambda_+ |\lambda_-|} & \sqrt{2 |\lambda_-|} \rmi \Delta & \sqrt{2 \lambda_+} \rmi \Delta \\ 0 & 0 & \sqrt{\Omega \lambda_+ |\lambda_-|} & \sqrt{2 |\lambda_-|} \rmi \Delta & \sqrt{2 \lambda_+} \rmi \Delta \\ 0 & \sqrt{\Omega \lambda_+ |\lambda_-|} & 0 & \sqrt{2 |\lambda_-|} \rmi \Delta & \sqrt{2 \lambda_+} \rmi \Delta \\ \sqrt{\Omega \lambda_+ |\lambda_-|} & 0 & 0 & \sqrt{2 |\lambda_-|} \rmi \Delta & \sqrt{2 \lambda_+} \rmi \Delta \end{pmatrix} \,,
\end{equation}
such that
\begin{equation}
  \mathbf{V}^{-1} \overline{\mathbf{M}} \mathbf{V} = \lambda \mathbb{I} - \begin{pmatrix} 0 & 0 & 0 & 0 & 0 \\ 0 & 0 & 0 & 0 & 0 \\ 0 & 0 & 0 & 0 & 0 \\ 0 & 0 & 0 & \lambda_+ & 0 \\ 0 & 0 & 0 & 0 & \lambda_- \end{pmatrix} \,.
\end{equation}

The matrix $\overline{\mathbf{G}}$ of Green functions is given by
\begin{equation}
  \overline{\mathbf{G}}(t) = \rme^{\lambda t} \, \mathbf{V} \begin{pmatrix} 1 & 0 & 0 & 0 & 0 \\ 0 & 1 & 0 & 0 & 0 \\ 0 & 0 & 1 & 0 & 0 \\ 0 & 0 & 0 & \rme^{-\lambda_+ t} & 0 \\ 0 & 0 & 0 & 0 & \rme^{-\lambda_- t} \end{pmatrix} \mathbf{V}^{-1} = \rme^{\lambda t} \begin{pmatrix} G_{\ti, \ti}(t) & G_{\ti, \ts}(t) & G_{\ti, \ts}(t) & G_{\ti, \ts}(t) & G_{\ti, \ts}(t) \\ G_{\ti, \ts}^*(t) & G_{\ts, \ts}(t) & G_{\ts, \ts'}(t) & G_{\ts, \ts'}(t) & G_{\ts, \ts'}(t) \\ G_{\ti, \ts}^*(t) & G_{\ts, \ts'}(t) & G_{\ts, \ts}(t) & G_{\ts, \ts'}(t) & G_{\ts, \ts'}(t) \\ G_{\ti, \ts}^*(t) & G_{\ts, \ts'}(t) & G_{\ts, \ts'}(t) & G_{\ts, \ts}(t) & G_{\ts, \ts'}(t) \\ G_{\ti, \ts}^*(t) & G_{\ts, \ts'}(t) & G_{\ts, \ts'}(t) & G_{\ts, \ts'}(t) & G_{\ts, \ts}(t) \end{pmatrix} \,,
\end{equation}
with matrix elements
\begin{subequations}\label{G}\begin{eqnarray}
  G_{\ti, \ti}(t) &=& \rme^{-\frac{1}{2} (\gamma_\ti - \gamma_\ts) t} \left( \cosh \frac{\Omega}{2} t - \frac{\gamma_\ti - \gamma_\ts}{\Omega} \sinh \frac{\Omega}{2} t \right) \,, \\
  G_{\ti, \ts}(t) &=& -2 \rmi \frac{\Delta}{\Omega} \rme^{-\frac{1}{2} (\gamma_\ti - \gamma_\ts) t} \sinh \frac{\Omega}{2} t \,, \\
  G_{\ts, \ts}(t) &=& \frac{3}{4} + \frac{\rme^{-\frac{1}{2} (\gamma_\ti - \gamma_\ts) t}}{4} \left( \cosh \frac{\Omega}{2} t + \frac{\gamma_\ti - \gamma_\ts}{\Omega} \sinh \frac{\Omega}{2} t \right) \,, \\
  G_{\ts, \ts'}(t) &=& G_{\ts, \ts}(t) - 1 \,.
\end{eqnarray}\end{subequations}
Convergence of these functions requires $\gamma_\ti + \gamma_\ts > \Omega$, i.e., $\gamma_\ti \gamma_\ts > 4 \Delta^2$.

Introduction of the uniform noise background $N_b$ allows for the evaluation of the idler and signal populations in the long-time limit.
This yields
\begin{eqnarray}
  N_{\ti, \ti}^\infty \equiv \lim_{t \to \infty} \langle p_\ti^\dagger(t) p_\ti(t) \rangle &=& \int_0^\infty 2 \rme^{-2 \gamma_\ts t} \Big\{ N_b \gamma_\ti G_{\ti, \ti}^2(\tau) + 4 (N_b + 1) \gamma_\ts |G_{\ti, \ts}(\tau)|^2 \Big\} \, \rmd\tau \nonumber\\
  &=& \frac{N_b \gamma_\ti}{\gamma_\ti + \gamma_\ts} \frac{\gamma_\ts (\gamma_\ti + \gamma_\ts) - 4 \Delta^2}{\gamma_\ti \gamma_\ts - 4 \Delta^2} + \frac{(N_b + 1) \gamma_\ts}{\gamma_\ti + \gamma_\ts} \frac{4 \Delta^2}{\gamma_\ti \gamma_\ts - 4 \Delta^2} \,,
\end{eqnarray}
\begin{equation}
  N_{\ts, \ts}^\infty \equiv \lim_{t \to \infty} \langle p_\ts^\dagger(t) p_\ts(t) \rangle = \frac{(N_b + 1) \gamma_\ti}{\gamma_\ti + \gamma_\ts} \frac{\Delta^2}{\gamma_\ti \gamma_\ts - 4 \Delta^2} + \frac{3 N_b}{4} + \frac{N_b \gamma_\ts}{4 (\gamma_\ti + \gamma_\ts)} \frac{\gamma_\ti (\gamma_\ti + \gamma_\ts) - 4 \Delta^2}{\gamma_\ti \gamma_\ts - 4 \Delta^2}\,,
\end{equation}
and the correlators become
\begin{equation}
  N_{\ts, \ts'}^\infty \equiv \lim_{t \to \infty} \langle P_\ts^\dagger(t) P_{i'}(t) \rangle = \frac{(N_b + 1) \gamma_\ti}{\gamma_\ti + \gamma_\ts} \frac{\Delta^2}{\gamma_\ti \gamma_\ts - 4 \Delta^2} - \frac{N_b}{4} + \frac{N_b \gamma_\ts}{4 (\gamma_\ti + \gamma_\ts)} \frac{\gamma_\ti (\gamma_\ti + \gamma_\ts) - 4 \Delta^2}{\gamma_\ti \gamma_\ts - 4 \Delta^2} \,,
\end{equation}
\begin{equation}
  N_{\ti, \ts}^\infty \equiv \lim_{t, t' \to \infty} \langle P_\ti^\dagger(t) P_\ts^\dagger(t') \rangle = \frac{\rmi (2 N_b + 1) \gamma_\ti \gamma_\ts \Delta}{(\gamma_\ti + \gamma_\ts) (\gamma_\ti \gamma_\ts - 4 \Delta^2)} \,.
\end{equation}
Finally, the tomographic reconstruction results in the state
\begin{equation}
  \rho = \frac{1}{4} \frac{N_{\ti, \ti}^\infty N_{\ts, \ts'}^\infty + |N_{\ti, \ts}^\infty|^2}{N_{\ti, \ti}^\infty N_{\ts, \ts}^\infty + |N_{\ti, \ts}^\infty|^2} \begin{pmatrix} 1 & 1 & 1 & 1 \\ 1 & 1 & 1 & 1 \\ 1 & 1 & 1 & 1 \\ 1 & 1 & 1 & 1 \end{pmatrix} + \frac{1}{4} \left( 1 - \frac{N_{\ti, \ti}^\infty N_{\ts, \ts'}^\infty + |N_{\ti, \ts}^\infty|^2}{N_{\ti, \ti}^\infty N_{\ts, \ts}^\infty + |N_{\ti, \ts}^\infty|^2} \right) \begin{pmatrix} 1 & 0 & 0 & 0 \\ 0 & 1 & 0 & 0 \\ 0 & 0 & 1 & 0 \\ 0 & 0 & 0 & 1 \end{pmatrix} \,,
\end{equation}
such that the value of $X$ in Eq.\ \eqref{rho} becomes
\begin{equation}
  X = \frac{N_{\ti, \ti}^\infty N_{\ts, \ts'}^\infty + |N_{\ti, \ts}^\infty|^2}{N_{\ti, \ti}^\infty N_{\ts, \ts}^\infty + |N_{\ti, \ts}^\infty|^2} \,.
\end{equation}

\end{widetext}

\bibliography{ref}

\end{document}